\documentclass[12pt]{article}
\usepackage{graphicx}

\newcommand\pubnumber{DPF2013-220}
\newcommand\pubdate{\today}

\def\napoli{Department  of Physics and Astronomy\\
University of Mississippi, Lewis Hall, University, MS, 38677, USA\\[5mm]
Department  of Physics, Faculty of Science\\
Ain Shams University, Cairo, 11566, Egypt}
\def\support{\footnote{This work supported  
in part by the National Science Foundation under Grant No.\ NSF PHY-1068052.}}

\def\Title#1{\begin{center} {\Large #1 } \end{center}}
\def\Author#1{\begin{center}{ \sc #1} \end{center}}
\def\Address#1{\begin{center}{ \it #1} \end{center}}

\newcommand\pubblock{\rightline{\begin{tabular}{l} \pubnumber\\
         \pubdate  \end{tabular}}}
\newenvironment{Abstract}{\begin{quotation}  }{\end{quotation}}
\newenvironment{Presented}{\begin{quotation} \begin{center} 
             PRESENTED AT\end{center}\bigskip 
      \begin{center}\begin{large}}{\end{large}\end{center} \end{quotation}}

\def\beq{\begin{equation}}
\def\eeq#1{\label{#1}\end{equation}}
\def\eeqn{\end{equation}}

\def\beqa{\begin{eqnarray}}
\def\eeqa#1{\label{#1}\end{eqnarray}}
\def\eeqan{\end{eqnarray}}

\let\bar=\overbar

\def\Dslash{\not{\hbox{\kern-4pt $D$}}}
\def\dslash{\not{\hbox{\kern-2pt $\del$}}}

\def\ee{e^+e^-}

\def\msb{{\bar{\ssstyle M \kern -1pt S}}}

\def\scat{  \nu_{\tau}+ N \to \tau^- + X}
\def\scatanti{ \bar{\nu}_{\tau}+ N \to \tau^+ + X}

\def\nutau{ \nu_{\tau}}
\def\numu{ \nu_{\mu}}

\def\taud{\tau^- \to \pi^- \nu_{\tau}}
\def\tauv{\tau^- \to \rho^- \nu_{\tau}}

\def\beq{\begin{equation}}
\def\lsim{\raise0.3ex\hbox{$\;<$\kern-0.75em\raise-1.1ex\hbox{$\sim\;$}}}
\def\gsim{\raise0.3ex\hbox{$\;>$\kern-0.75em\raise-1.1ex\hbox{$\sim\;$}}}
\def\eeq{\end{equation}}
\def\be{\begin{equation}}
\def\ee{\end{equation}}
\def\bea{\begin{eqnarray}}
\def\eea{\end{eqnarray}}

\begin{document}

\begin{titlepage}
\pubblock

\vfill
\Title{Tau neutrino as a probe of nonstandard interactions via charged Higgs and $W'$ contribution}
\vfill
\Author{Ahmed Rashed\support}
\Address{\napoli}
\vfill
\begin{Abstract}
We discuss the impact of the presence of a charged Higgs and a $W'$ gauge boson on the tau-neutrino nucleon scattering $\scat$ and $\scatanti$. We show the effect of the new physics on the three subprocesses quasielastic, $\Delta$-resonance, and deep inelastic scattering. The measurement of the atmospheric and reactor mixing angles $\theta_{23}$ and $\theta_{13}$, respectively, relies on the standard model cross section of the above processes if they have been measured in the appearance channels $\nu_\mu \rightarrow \nu_\tau$ and $\bar{\nu}_e \rightarrow \bar{\nu}_\tau$ ($\nu_e \rightarrow \nu_\tau$), respectively. Consideration of the new physics contributions to those reactions modifies the measured mixing angles, assuming the standard model cross section. We include form factor effects in the new physics calculations and find the deviations of the mixing angles which can be significant and can depend on the energy of the neutrino.
\end{Abstract}
\vfill
\begin{Presented}
DPF 2013\\
The Meeting of the American Physical Society\\
Division of Particles and Fields\\
Santa Cruz, California, August 13--17, 2013\\
\end{Presented}
\vfill
\end{titlepage}
\def\thefootnote{\fnsymbol{footnote}}
\setcounter{footnote}{0}

The effects of nonstandard interaction (NSI) on neutrino oscillation have been widely studied \cite{Wolfenstein:1977ue}. It has been established that NSI cannot be an explanation for the standard oscillation phenomena, but it may be present as a subleading effect. General bounds on the NSI parameters have been discussed in the literatures \cite{davidson}. The NSI impact has been studied on different themes in neutrino phenomenology \cite{impact}. Often in the analysis of NSI, hadronization effects of the quarks via form factors are not included. In Ref.~\cite{Rashed:2012bd, Rashed:2013dba}, the results show that the form factors play an important role in the energy dependence of the NSI effects.  Many NSI involve flavor changing neutral current or charged current lepton flavor violating processes. Here we consider charged current interactions involving contributions from a charged Higgs and a $W'$ gauge boson.  In neutrino experiments, to measure the mixing angle the neutrino-nucleus interaction is assumed to be SM-like. If there is a charged Higgs or a $W'$ contribution to this interaction, then there will be an error in the extracted mixing angle. We will calculate the error in the extracted mixing angle. Constraints on the new couplings come from the hadronic $\tau$ decays. We will consider constraints from the decays $\taud$ and $\tauv$ \cite{Rashed:2012bd, Rashed:2013dba}.

There are several reasons to consider NSI involving the $( \nu_\tau, \tau)$ sector. First, the third generation may be more sensitive to new physics effects because of their larger masses. As an example, in certain versions of the two Higgs doublet models (2HDM) the couplings of the new Higgs bosons are proportional to the masses, and so new physics effects are more pronounced for the third generation. Second, the constraints on NP involving the third generation leptons are somewhat weaker, allowing for larger new physics effects. Interestingly, the branching ratio of $B$ decays to $\tau$ final states shows some tension with the SM  predictions \cite{belletau, Lees:2012xj} and this could indicate NP, possibly in the scalar or gauge boson sector \cite{Datta:2012qk}. Some examples of work that deals with NSI at the detector, though not necessarily involving the third family leptons, can be found in Refs.~\cite{nir, biggio}.

The process $\scat$ will impact the measurement of the oscillation probability for the $\numu \to \nutau$ transition and hence the extraction of the mixing angle $\theta_{23}$. The measurement of the atmospheric mixing angle $\theta_{23}$ relies on the following relationship \cite{relationship}:
\beq
N(\nu_\tau) = P(\nu_\mu \rightarrow \nu_\tau) \times \Phi (\nu_\mu)\times \sigma_{{\rm SM}}(\nu_\tau)\,,
\label{eq-1}
\eeq
where $N(\nu_\tau)$ is the number of observed events, $\Phi (\nu_\mu)$ is the flux of muon neutrinos at the detector,   $\sigma^{{\rm SM}}(\nu_\tau)$  is the total cross section of tau neutrino interactions with nucleons in the SM at the detector, and $P(\nu_\mu \rightarrow \nu_\tau)$ is the probability for the flavor transition $\numu \to \nutau$. This probability is a function of $(E,\; L,\; \Delta m_{ij}^2,\; \theta_{ij})$ with $i,j=1,2,3$, where $\Delta m_{ij}^2$ is the squared-mass difference, $\theta_{ij}$ is the mixing angle, $E$ is the energy of neutrinos, and $L$ is the distance traveled by neutrinos. The dominant term of the probability is
\beq
P(\nu_\mu \rightarrow \nu_\tau) \approx \sin^2 2\theta_{23} \cos^4 \theta_{13} \sin^2 (\Delta m^2_{23} L/4E).
\eeq
In the presence of NP, Eq.~\ref{eq-1} is modified as
\beq
N (\nu_\tau) = P(\nu_\mu \rightarrow \nu_\tau) \times \Phi (\nu_\mu)\times \sigma_{{\rm tot}}(\nu_\tau),
\label{eq-2}
\eeq
with $\sigma_{{\rm tot}}(\nu_\tau)=\sigma_{{\rm SM}}(\nu_\tau)+\sigma_{{\rm NP}}(\nu_\tau)$, where $\sigma_{{\rm NP}}(\nu_\tau)$ refers to the additional terms of the SM contribution towards the total cross section. Hence, $\sigma_{{\rm NP}}(\nu_\tau)$  includes  contributions from both the SM and NP interference amplitudes, and the pure NP amplitude. { }From Eqs.~(\ref{eq-1}, \ref{eq-2}), assuming $\theta_{13}$ to be small,{\footnote{The presence of NP impacts the extraction of the combination $\sin^2 2 \theta_{23}\cos^4 \theta_{13}$. The NP changes the extracted value of $\theta_{23}$ as well as $\theta_{13}$. But we fix the value of $\theta_{13}$ as an input at this point.}}
\bea
\label{modineq3}
\sin^2{2( \theta_{23})} &=& \sin^2{2( \theta_{23})_{SM}} \frac{1}{1 + r_{23}}\,,
\eea
where $\theta_{23}= (\theta_{23})_{SM} +\delta_{23}$ is the actual atmospheric mixing angle, whereas $(\theta_{23})_{SM}$ is the extracted mixing angle assuming the SM $\nu_\tau$ scattering cross section. Assuming negligible new physics effects in the $\mu-N$ interaction, the actual mixing angle $\theta_{23}$ is the same as the mixing angle extracted from the survival probability $P(\nu_\mu \rightarrow \nu_\mu)$ measurement. We will take the best-fit value for the mixing angle to be given by $ \theta_{23} = 42.8^\circ$ \cite{GonzalezGarcia:2010er}. In other words, the presence of new physics in a $\nu_\tau$-nucleon scattering will result in the mixing angle, extracted from a $\nu_\tau$ appearance experiment, being different than the mixing angle from $\nu_\mu$ survival probability measurements. The relationship between the ratio of the NP contribution to the SM cross section $ r_{23} = \sigma_{NP}(\nu_\tau) / \sigma_{SM}(\nu_\tau)$ and $\delta_{23}$ can be expressed in a model-independent form as
\bea
\label{modineq4}
 r_{23} &=&\Big[ \frac{\sin{2( \theta_{23})_{SM}}}{\sin{2( (\theta_{23})_{SM} + \delta_{23}) }}\Big]^2-1\,.
\eea

\begin{figure}[htb]
\centering
\includegraphics[width=6.70cm]{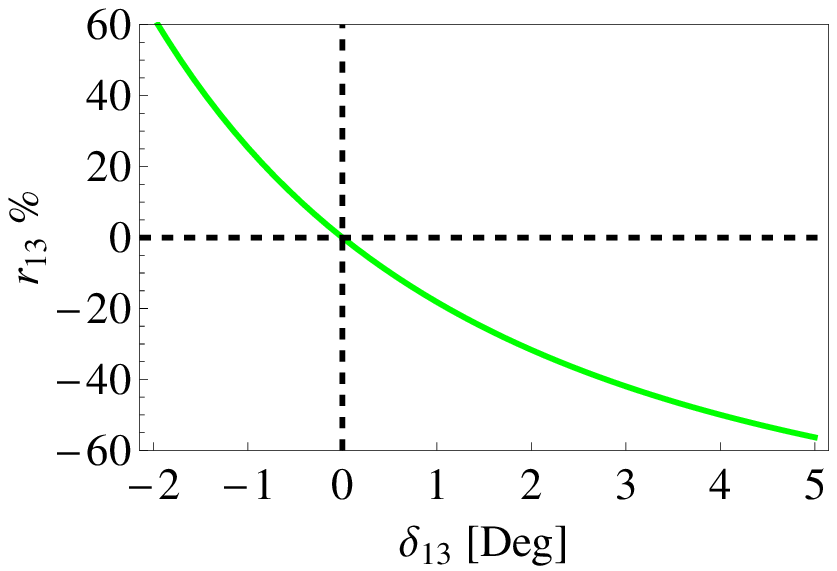}
\includegraphics[width=6.40cm]{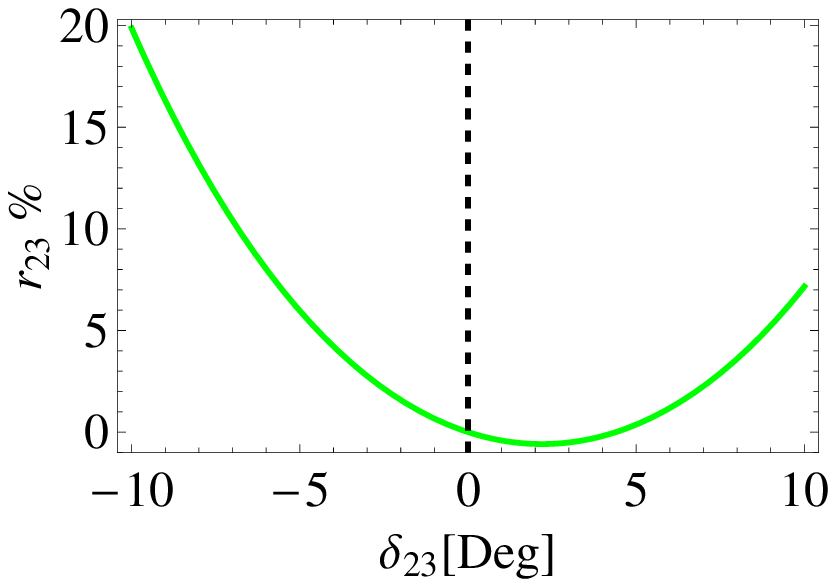}
\caption{Correlation plot for $r_{23} = \sigma_{NP}(\nu_\tau) / \sigma_{SM}(\nu_\tau) \%$ versus $\delta_{23} [Deg]$, and $r_{13} = \sigma_{NP}(\bar{\nu}_\tau) / \sigma_{SM}(\bar{\nu}_\tau) \%$ versus $\delta_{13} [Deg]$.}
\label{r23del23fig}
\end{figure}

The reactor mixing angle $\theta_{13}$ can be determined from the oscillation probability of the appearance channel $\bar{\nu}_e \rightarrow \bar{\nu}_\tau$ ($\nu_e \rightarrow \nu_\tau$). In this case the effect of NP contributions to the process $\scatanti(\scat)$ is pertinent. The best-fit value for the mixing angle to be given by $ \theta_{13} = 9.1^\circ$ \cite{Tortola:2012te}. Many neutrino mixing models have expected non-zero value for $\theta_{13}$ \cite{models}. The relationship, e.g., used in measuring $\theta_{13}$ will be given as
\beq
N(\bar{\nu}_\tau) = P(\bar{\nu}_e \rightarrow \bar{\nu}_\tau) \times \Phi (\bar{\nu}_e)\times \sigma_{{\rm tot}}(\bar{\nu}_\tau)\,,
\label{eq-111}
\eeq
where \cite{Donini:2002rm, Upadhyay:2011aj, Huber:2006wb}
\beq
P(\bar{\nu}_e \rightarrow \bar{\nu}_\tau) \approx \sin^2 2\theta_{13} \cos^2 \theta_{23} \sin^2 (\Delta m^2_{13} L/4E).
\eeq
Thus the relationship  between the ratio of the NP contribution to the SM cross section $ r_{13} = \sigma_{NP}(\bar{\nu}_\tau) / \sigma_{SM}(\bar{\nu}_\tau)$ and $\delta_{13}$ can be obtained in a model-independent form as
\bea
\label{modineq444}
 r_{13} &=&\Big[ \frac{\sin{2( \theta_{13})_{SM}}}{\sin{2( (\theta_{13})_{SM} + \delta_{13}) }}\Big]^2-1\,.
\eea
In Fig.~\ref{r23del23fig} we show the correlation between  $ r_{23(13)} \%$ and $\delta_{23(13)}$ [Deg]. One can see that  $\delta_{23}  \sim - 5^\circ $  requires $ r_{23} \sim 5\% $. But $\delta_{13}  \sim - 1^\circ $  requires $ r_{13} \sim 25\%$. 

The  coupling of charged Higgs boson ($H^{\pm}$) interactions to a SM fermion in the 2HDM II is \cite{Diaz:2002tp}
\beq
\label{HiggsLag1}
\mathcal{L} = \frac{g}{\sqrt{2}M_W} \sum_{ij} \Big[m_{u_i}  \cot{\beta} ~  \bar{u}_{i} V_{ij} P_{L,R} d_{j} + m_{d_j} \tan{\beta} ~  \bar{u}_{i} V_{ij} P_{R,L} d_{j}  + m_{l_j} \tan{\beta}~ \bar{\nu}_{i} P_{R,L} l_{j} \Big] H^{\pm},
\eeq
 where $P_{L,R}= {( 1 \mp \gamma^5) / 2}$, and $\tan \beta$ is the ratio between the two vacuum expectation values (vev's) of the two Higgs doublets, and
\bea
\label{2HDMcoup}
g^{u_i d_j}_S &=&  \left (\frac{m_{d_j} \tan{\beta} + m_{u_i} \cot{\beta}}{M_W} \right), \nonumber\\
g^{u_i d_j}_P &=&  \left (\frac{m_{d_j} \tan{\beta} - m_{u_i} \cot{\beta}}{M_W} \right),\nonumber\\
g^{\nu_i l_j}_S &=& g^{\nu_i l_j}_P = \frac{m_{l_j} \tan{\beta}}{M_W}.
\eea
The lowest dimension effective Lagrangian of $W^\prime$ interactions to the SM fermions has the form 
\bea
{\cal{L}} &=& \frac{g}{\sqrt{2}}V_{ f^\prime f} \bar{f}^\prime \gamma^\mu( g^{f^\prime f}_L P_L +  g^{f^\prime f}_R P_R) f W^\prime_\mu + ~h.c.,
\label{wprime}
\eea
where  $f^\prime$ and $f$ refer to the fermions and $g^{f^\prime f}_{L,R}$ are the left- and the right-handed couplings of the $W^\prime$. {}For a SM-like $W^\prime$ boson, $g^{f^\prime f}_{L}=1 $ and $g^{f^\prime f}_{R}= 0$. We will assume $g^{f^\prime f}_{L,R}$ to be real. 

\begin{figure}[htb!]
\centering
 \includegraphics[width=5cm]{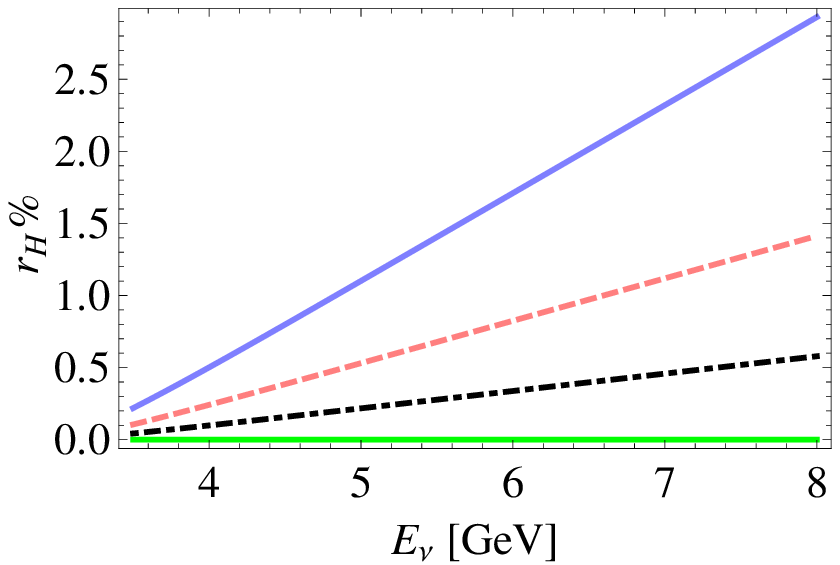}~~~
 \includegraphics[width=5cm]{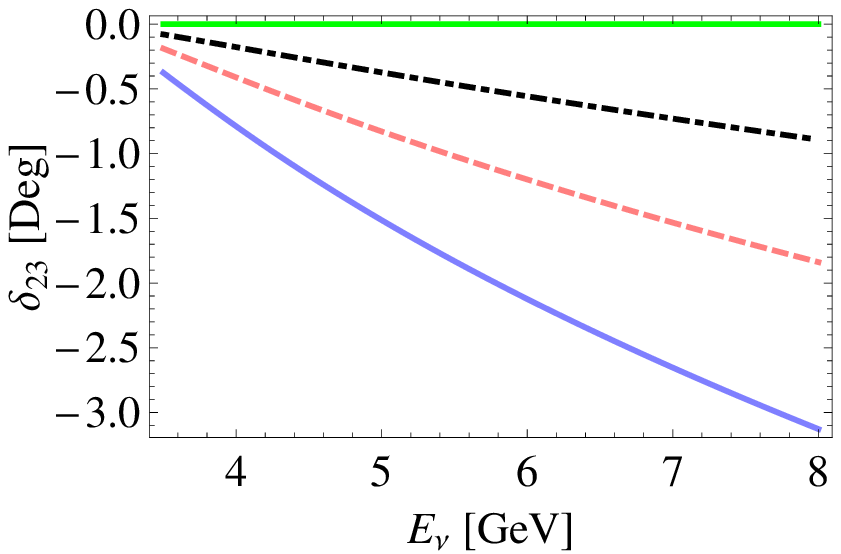}~~~
 \includegraphics[width=5cm]{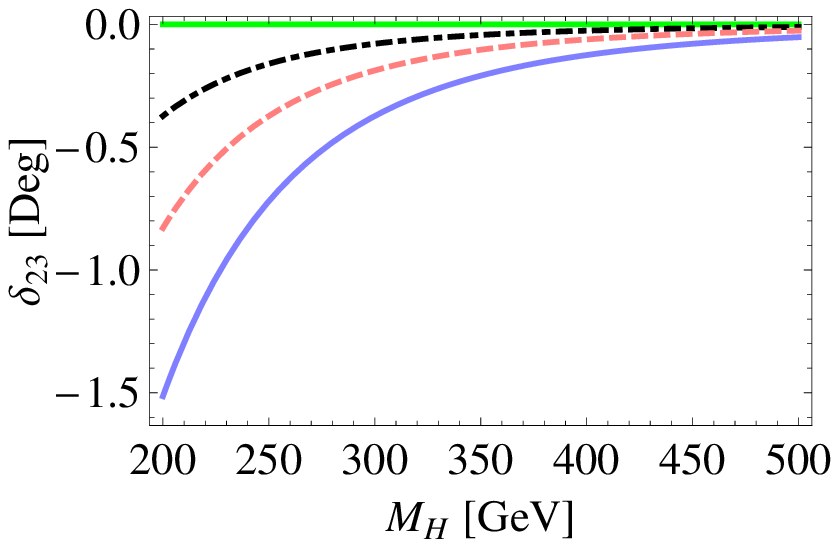}~~~
\caption{Quasi-elastic $(H^+)$: Variation of $r_H^{23} \%$ with $E_\nu$ and variation of $\delta_{23}$ with $M_H$  and $E_\nu$. The green line corresponds to the SM prediction.  The black (dotdashed), pink (dashed), and blue (solid) lines correspond to $\tan{\beta} = 40, 50, 60$. The right figure is evaluated  at $E_\nu = 5$ GeV, while the left figures are evaluated at $M_{H} = 200$ GeV. Here, we use the best-fit value $ \theta_{23} = 42.8^\circ$ \cite{GonzalezGarcia:2010er}.}
\label{delHvsMHplot}
\end{figure}

\begin{figure}[htb!]
\centering
 \includegraphics[width=5cm]{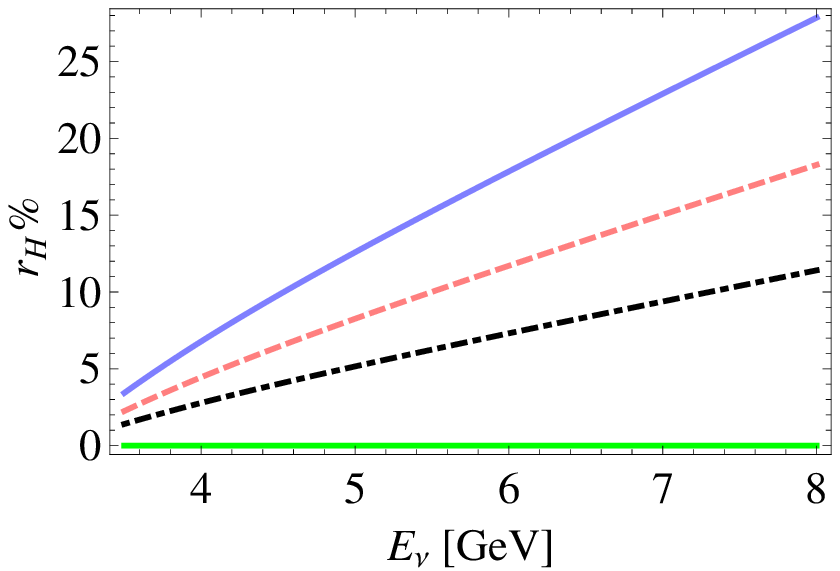}~~~
 \includegraphics[width=5cm]{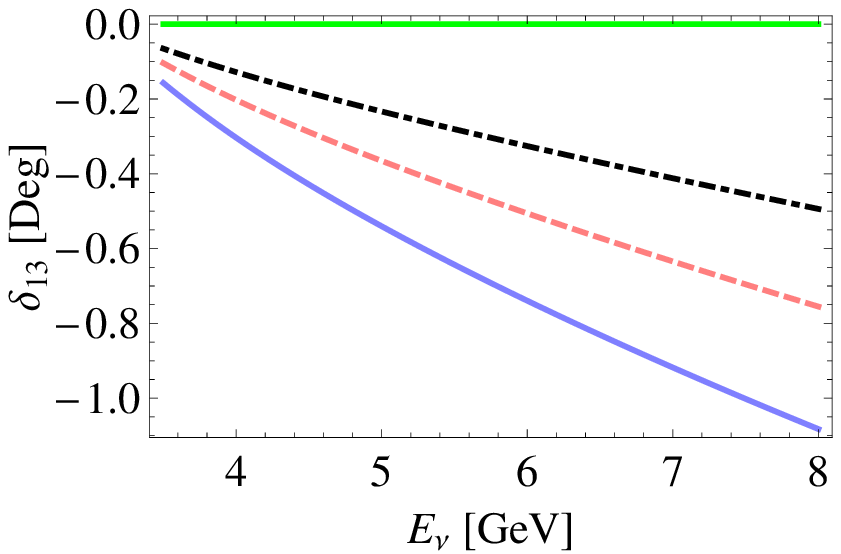}~~~
 \includegraphics[width=5cm]{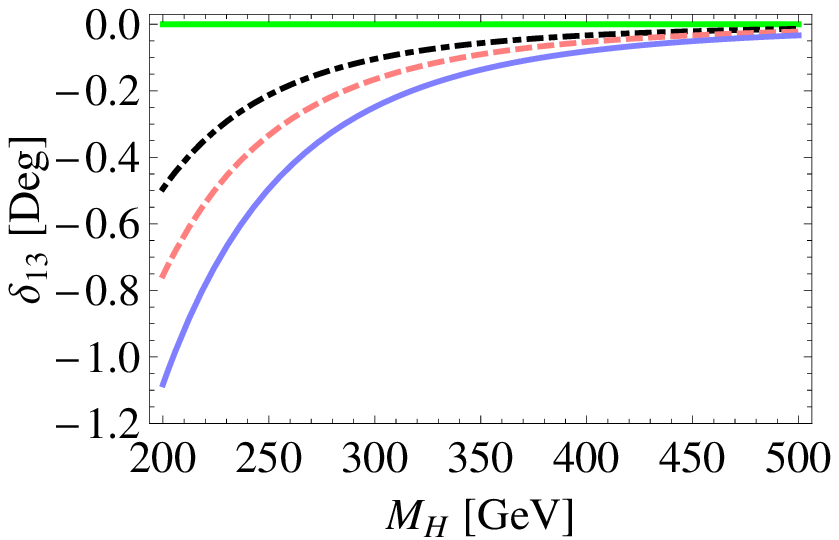}~~~
\caption{Quasi-elastic $(H^+)$: Variation of $r_H^{13} \%$ with $E_\nu$ and the variation of $\delta_{13}$ with $M_H$  and $E_\nu$. The green line corresponds to the SM prediction.  The black (dotdashed), pink (dashed), and blue (solid) lines correspond to $\tan{\beta} = 80, 90, 100$. The right figure is evaluated  at $E_\nu = 8$ GeV, while the left figures are evaluated at $M_{H} = 200$ GeV. Here, we use the inverted hierarchy value $ \theta_{13} = 9.1^\circ$ \cite{Tortola:2012te}.}
\label{delHvsMHplot000}
\end{figure}

\begin{figure}[h!]
\centering
 \includegraphics[width=5cm]{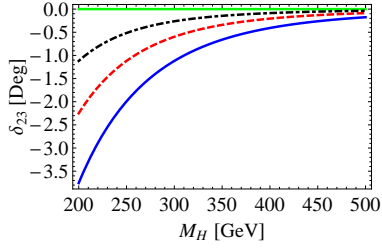}
\caption{Resonance ($H$): The figures illustrate variation of $\delta_{23}$ with $M_H$. The green line corresponds to the SM prediction. The black (dotdashed), red (dashed), and blue (solid) lines correspond to $\tan{\beta} = 40, 50, 60$. Here, we use the best-fit value $  \theta_{23} = 42.8^\circ$ \cite{GonzalezGarcia:2010er}. We take into account the atmospheric neutrino flux for Kamioka where the Super-Kamiokande experiment locates \cite{Honda:2011nf}.}
\label{Delta-RES-Flux-H}
\end{figure}

\begin{figure}[h!]
\centering
 \includegraphics[width=5cm]{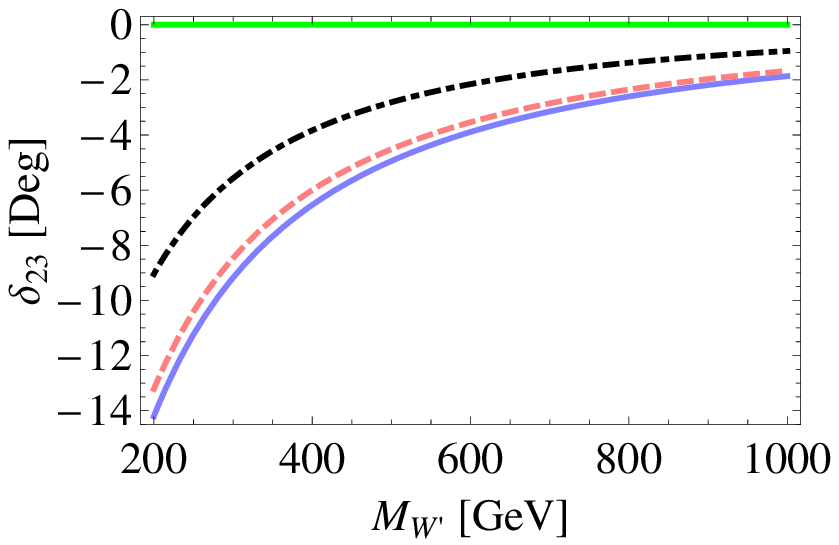}
\caption{Resonance ($W^\prime$): The figure illustrates the deviation $\delta_{23}$ with the $W^\prime$ mass $M_{W'}$ when both left and right-handed  $W^\prime$ couplings are present. The lines show predictions for some representative values of the $W^\prime$ couplings $(g^{\tau \nu_\tau}_L, g^{ud}_L, g^{ud}_R)$. The green line (solid, upper) corresponds to the SM prediction. The blue line (solid, lower) corresponds to (-0.94 ,  -1.13 , -0.85). Here, we use the best-fit value $  \theta_{23} = 42.8^\circ$ \cite{GonzalezGarcia:2010er}. We take into account the atmospheric neutrino flux for Kamioka where the Super-Kamiokande experiment locates \cite{Honda:2011nf}.}
\label{Delta-RES-Flux-Wp}
\end{figure}

\begin{figure}[h!]
\centering
 \includegraphics[width=5cm]{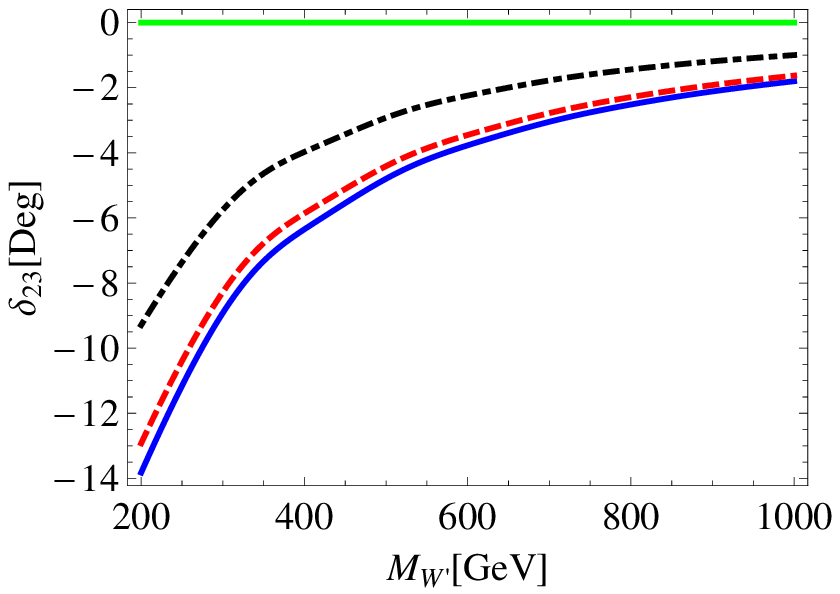}
\caption{DIS ($W'$): The figure illustrates the deviation $\delta_{23}$ with the $W^\prime$ mass $M_{W'}$ when both left and right-handed  $W^\prime$ couplings are present. The lines show predictions for some representative values of the $W^\prime$ couplings $(g^{\tau \nu_\tau}_L, g^{ud}_L, g^{ud}_R)$. The green line (solid, upper) corresponds to the SM prediction. The blue line (solid, lower) corresponds to (-0.94 ,  -1.13 , -0.85). Here, we use the best-fit value $  \theta_{23} = 42.8^\circ$ \cite{GonzalezGarcia:2010er}. We take into account the atmospheric neutrino flux for Kamioka where the Super-Kamiokande experiment locates \cite{Honda:2011nf}.}
\label{DIS-Flux-Wp}
\end{figure}

In Figs.~(\ref{delHvsMHplot}, \ref{delHvsMHplot000}), we show the deviation of the mixing angles $\theta_{23}$ and $\theta_{13}$ due to the contribution of the charged Higgs to the tau-neutrino cross section. 
The deviations $\delta_{23}$ and $\delta_{13}$ are negative, as there is no interference with the SM; hence, the cross section for $\scat$ and $\scatanti$ are always larger than the SM cross section. This means that, if the actual $\theta_{23}$ is close to maximal, then experiments should measure $\theta_{23}$ larger than the maximal value in the presence of a charged Higgs contribution. 
In the $\Delta$-RES and DIS cases, their effect has been introduced with considering the flux of the incoming muon-neutrinos and integrating over the possible atmospheric neutrino energy. In Figs.~(\ref{Delta-RES-Flux-H}, \ref{Delta-RES-Flux-Wp}, \ref{DIS-Flux-Wp}), we show the deviation $\delta_{23}$ with including the flux effect in the $\Delta$-RES and DIS cases.

We calculate the number of events in the DIS $W'$ model. We compare between the number of events of the atmospheric neutrinos in the SM $N_{\rm SM}=30.66\pm 3.37$ and in the NP $N_{NSI}=41.49$ for $M_{W'}=200$ GeV. This shows that the $N_{NSI}$ falls beyond the uncertainty of $N_{\rm SM}$ and so the NSI is potentially detectable.

\vspace{5mm}



\end{document}